\documentclass{article}
\usepackage{amsmath}
\advance\topmargin by -20mm \advance\textheight by 45mm \textwidth=165mm
\newcommand{\ket}[1]{\ensuremath{\left|#1\right>}}

\newcommand{\cuo}{\ensuremath{\text{CuO}_2}}
\newcommand{\ie}{\textit{i.e.},~}

\renewcommand{\r}{\ensuremath{\mathbf{r}}}
\renewcommand{\k}{\ensuremath{\mathbf{k}}}
\newcommand{\ax}{\ensuremath{\mathbf{a}_x}}
\newcommand{\ay}{\ensuremath{\mathbf{a}_y}}

\begin{document}
\title{Simple theory of extremely overdoped HTS}
\author{A.F. Andreev\footnote{E-mail: andreev@kapitza.ras.ru}}
\date{}
\maketitle

\begin{center}
\textit{Kapitza Institute for Physical Problems, Russian Academy of Sciences,\\
Kosygin Str. 2, Moscow, 119334 Russia,\\
Low Temperature Laboratory, Helsinki University of Technology,\\
FIN-02015 HUT, Finland}
\end{center}

\begin{abstract}
We demonstrate the existence of a simple physical picture of
superconductivity for extremely overdoped \cuo\ planes. It possesses
all characteristic features of HTS, such as a high
superconducting transition temperature, the $d_{x^2 - y^2}$ symmetry
of order parameter, and the coexistence of a single electron Fermi
surface and a pseudogap in the normal state. Values of pseudogap are
calculated for different doping levels. An orbital paramagnetism of
preformed pairs is predicted.
\end{abstract}


\section{Introduction}
In this work (see also earlier Letter \cite{lit0}), we demonstrate that, in the phase diagram of cuprate HTS,
a small region exists in which the characteristic features of HTS can
be easily understood on the bases of a simple theory. The above
characteristic features are such as a high superconducting transition
temperature, the $d_{x^2 - y^2}$ symmetry of order parameter (see \cite{lit1}), and the
coexistence in the normal state of a single electron Fermi surface and a
pseudogap \cite{lit2}. The latter phenomenon is usually attributed to the presence of
preformed (\ie in the normal state) electron pairs (in particular,
bipolarons [4-8]).

The aforementioned small region in the phase diagram is situated in the
vicinity of the maximal hole-doping level $x = x_c$ compatible with
superconductivity. The superconducting transition temperature $T_c$ is
zero for $x \ge x_c$, so it is low in our region near $x = x_c$. However,
$T_c$ increases with decreasing $x$ for $x < x_c$ in such a way that at
the boundary of the region (i.e., for $x_c - x \sim 1$) it is quite high.

Two features of our small region are important to make a simple
physical picture possible. These are relatively low $T_c$ and the clear
nature of the normal state as mostly the conventional Fermi liquid.

We calculate the pseudogap. With increasing $x$, the pseudogap decreases for
$x < x_c$. As well as $T_c$, the pseudogap disappears at $x = x_c$.
However, for larger doping levels $x > x_c$, it appears again.

As a new prediction, we show the existence of an unusual orbital paramagnetism
of the preformed (singlet) pairs, which probably can be experimentally
separated from the Pauli spin paramagnetism of single electrons and the Landau
diamagnetism of single electrons and pairs.

\section{Pair quasiparticles}
The key point is the existence of very mobile
pair quasiparticles in crystals in conditions of the
tight-binding, \ie if the energy of
electron-electron interaction at a distance on the order of atomic
spacing considerably exceeds the electron-tunneling amplitude to
neighboring lattice cites. Quasiparticles of this type were studied
earlier \cite{lit8} in helium quantum crystals and more recently by Alexandrov and
Kornilovitch \cite{lit6} as a model of bipolarons in HTS (see also \cite{lit9}).

\begin{figure}[h]
\begin{center}
\begin{picture}(180,180)(2,689)
\put(152,832){\makebox(0,0)[lb]{\smash{{$\bullet\ 3$%
}}}}
\put(152,795){\makebox(0,0)[lb]{\smash{{$\times$%
}}}}
\put(152,726){\makebox(0,0)[lb]{\smash{{$\times$%
}}}}
\put(  2,832){\makebox(0,0)[lb]{\smash{{$\bullet\ 7$%
}}}}
\put( 40,832){\makebox(0,0)[lb]{\smash{{$\times$%
}}}}
\put( 77,832){\makebox(0,0)[lb]{\smash{{$\bullet\ 4$%
}}}}
\put( 77,689){\makebox(0,0)[lb]{\smash{{$\bullet\ 6$%
}}}}
\put( 40,689){\makebox(0,0)[lb]{\smash{{$\times$%
}}}}
\put(  2,762){\makebox(0,0)[lb]{\smash{{$\bullet\ $%
}}}}
\put( 40,762){\makebox(0,0)[lb]{\smash{{$\times$%
}}}}
\put( 77,762){\makebox(0,0)[lb]{\smash{{$\bullet\ 1$%
}}}}
\put( 77,726){\makebox(0,0)[lb]{\smash{{$\times$%
}}}}
\put(  2,726){\makebox(0,0)[lb]{\smash{{$\times$%
}}}}
\put(  2,689){\makebox(0,0)[lb]{\smash{{$\bullet\ $%
}}}}
\put(  2,795){\makebox(0,0)[lb]{\smash{{$\times$%
}}}}
\put(114,832){\makebox(0,0)[lb]{\smash{{$\times$%
}}}}
\put( 77,795){\makebox(0,0)[lb]{\smash{{$\times$%
}}}}
\put(114,762){\makebox(0,0)[lb]{\smash{{$\times$%
}}}}
\put(152,762){\makebox(0,0)[lb]{\smash{{$\bullet\ 2$%
}}}}
\put(114,689){\makebox(0,0)[lb]{\smash{{$\times$%
}}}}
\put(152,689){\makebox(0,0)[lb]{\smash{{$\bullet\ 5$%
}}}}
\end{picture}%
\end{center}
\caption{\cuo\  plane: $(\bullet)$ Cu atoms and $(\times)$ O atoms.}
\end{figure}

Let us
consider two electrons localized at neighboring (1 and 2 in the figure)
copper atoms (to be more precise, in unit cells containing these atoms)
forming a square lattice in the \cuo\  plane. The electron tunneling from
2 to 4 or 6 does not change the energy of the system in view of the
crystal lattice symmetry. The same is true for the electron tunneling
from 1 to 3 or 5. Owing to this type of transitions, an electron pair
can move as a whole over the entire plane, since the $2 \to 4$ transition can
be followed by the transition $1 \to 7$ or $1 \to 3$, and so on. Since the
transitions do not change the energy of the system, the motion is fully
coherent. An electron pair behaves as a delocalized Bose quasiparticle.

To calculate the quasiparticle spectrum, we consider the localized
states of a pair,
\begin{equation}
\ket{\r,\r',\alpha\beta}
=c^+_{\r\alpha}
 c^+_{\r'\beta}
 \ket{0},
\end{equation}
where
$c^+_{\r\alpha}$
are the electron creation operators at point $\r$ with spin
projection $\alpha=\uparrow,\downarrow$ and \ket{0} is the electron vacuum.

The effective tunneling
Hamiltonian $H_{eff}$ is defined by the matrix elements of the operator
\begin{equation}
H=t\sum\limits_{\r\r'\alpha}
     c^+_{\r'\alpha} 
     c_{\r\alpha},
\end{equation}
which correspond to the transitions of one of the electrons to copper
atoms that are next-to-nearest neighbors of the initial atom, in such a
way that the energy of the system of two electrons remains unchanged.
Here, $t$ is the tunneling amplitude, which is known to be positive (see
\cite{lit1}, p. 1004).

Let $\mathbf{a}_n$ ($n=x,y$)
be the square-lattice periods directed
from point 1 to point 2 and from point 1 to point 4, respectively. We
have
\begin{multline}
H_{eff} \ket{\r,\r+\ax,\alpha\beta}=
      t(
        \ket{\r+\ax+\ay,\r+\ax,\alpha\beta} + 
        \ket{\r+\ax-\ay,\r+\ax,\alpha\beta} + \\
        \ket{\r,\r+\ay,\alpha\beta} + 
        \ket{\r,\r-\ay,\alpha\beta})=
      t(
        -\ket{\r+\ax,\r+\ax+\ay,\beta\alpha} + 
        \ket{\r+\ax-\ay,\r+\ax,\alpha\beta} + \\
        \ket{\r,\r+\ay,\alpha\beta} - 
        \ket{\r-\ay,\r,\beta\alpha}),
\end{multline}
where we used the antisymmetry of quantities (1) with respect
to arguments
$(\r,\alpha)$ and
$(\r',\beta)$. Analogously,
\begin{multline}
H_{eff} \ket{\r,\r+\ay,\alpha\beta}=
      t(
        -\ket{\r+\ay,\r+\ax+\ay,\beta\alpha} + 
        \ket{\r-\ax+\ay,\r+\ay,\alpha\beta} + \\
        \ket{\r,\r+\ax,\alpha\beta} - 
        \ket{\r-\ax,\r,\beta\alpha}).
\end{multline}

The complete set of localized states of an electron
pair is determined by the state vectors
\begin{equation}
\ket{\r,n,\alpha\beta} \equiv \ket{\r,\r+\mathbf{a}_{n},\alpha\beta},
\end{equation}
where \r\ labels unit cells of the square lattice.

The problem obviously
splits into two independent problems for singlet and triplet pairs that
are characterized by quantities (5), respectively, antisymmetric and
symmetric about the spin indices $\alpha,\beta$. Assuming that the required
stationary states of a pair are superpositions of localized states,
\begin{equation}
\sum\limits_{\r,n}\psi_{\alpha\beta}^{(n)} e^{i\k\r} \ket{\r,n,\alpha\beta}
\end{equation}
with coefficients $\psi_{\alpha\beta}^{(n)}$ independent of \r\ (this
corresponds to a definite quasimomentum \k), we obtain
\begin{gather}
(E(\k)-\epsilon_0)\psi^{(x)}=t\psi^{(y)}(1 \pm e^{-i\kappa_x})(1 \pm e^{i\kappa_y}),\notag\\
(E(\k)-\epsilon_0)\psi^{(y)}=t\psi^{(x)}(1 \pm e^{i\kappa_x})(1 \pm e^{-i\kappa_y}),
\end{gather}
where the upper or lower sign corresponds to a singlet or triplet state,
respectively. The conditions for the existence of a nontrivial solution
$\psi^{(x)}$, $\psi^{(y)}$ to system (7) defines the energy $E(\k)$ of a
pair quasiparticle. Here $\epsilon_0$ is the energy of the initial
localized state; $\kappa_x=\k\ax$ and $\kappa_y=\k\ay$. Everywhere in
formulas (7), we omitted identical spin indices $\alpha\beta$.

The minimal energy $\epsilon_m=\min E(\k)=\epsilon_0-4t$ of a singlet
pair is attained for $\kappa_x=\kappa_y=0$. The same minimal energy of a
triplet pair is attained for a nonzero quasimomentum
$\kappa_x=\kappa_y=\pi$. This degeneracy is removed by taking into
account the electron exchange in the initial localized pair. It is well
known that this exchange is of an antiferromagnetic nature and, hence,
singlet pairs possess the minimal energy.

Thus, solitary Bose quasiparticles can exist in the \cuo\  plane; these
particles are characterized by a doubled electric charge and by zero
momentum and spin in the ground state. It can readily be seen from Eqs.
(7) that the effective mass of quasiparticles is $m=\hbar^2/ta^2$, where
$a=|\ax|=|\ay|$. In addition, quasiparticles possess a specific quantum
number $n=x,y$, which determines the orientation of a two-electron
``dumbbell''. Substituting $E(\k)=\epsilon_m$ and $\k=0$ into Eqs. (7),
we obtain $\psi^{(x)}=-\psi^{(y)}$ in the ground state. Since
orientations $n = x$ and $n = y$ are transformed into each other upon
the lattice rotation through an angle of $\pi/2$ and upon the reflection
in the diagonal plane passing through points 1 and 3 in the figure, the
ground-state wave function $\psi\equiv\psi^{(x)}=-\psi^{(y)}$ of
quasiparticles transforms in accordance with the nontrivial
1D representation
(usually denoted by $d_{x^2 - y^2}$) of the symmetry group of
\cuo\  plane (see \cite{lit1}).

\section{Superconductivity}
We further assume that all other two-, three-, etc.,
electron configurations localized at distances on the order of atomic
spacing are energetically disadvantageous as compared to the pair
configuration considered above. In addition, we assume that electrons
are repulsed at large distances such that the electron-electron
interaction energy is on the order of the one-electron tunneling
amplitude. Under these conditions, only single electron Fermi particles and
the pair Bose particles considered above play a significant role.

Finally, let us assume that the minimal energy $\epsilon_m$ of pair
quasiparticles is such that $\epsilon_m /2$ is within a single electron
energy band. We note the following. In conditions of the tight-binding,
there are two different situations in which $\epsilon_m /2$ can be
within a single electron energy band. First, if single electrons and
electrons in pairs correspond to the same energy band, single electron
tunneling amplitude should be of the order of the electron-electron
interaction energy in the pairs, while one-electron tunneling amplitude
$t$ in the pairs, introduced in Section 2, should be much smaller than
the interaction. The last condition which is the condition of
applicability of the procedure used in Section 2, can be result of the
large polaron effect in the pairs. In the second situation, single
electrons and electrons in pairs correspond to different bands. Both
one-electron amplitudes can be of the same order in this case. The
analysis carried out by Alexandrov and Kornilovitch
in \cite{lit6} shows that
the conditions formulated above are likely to be realistic.

Let us trace the change of the state of the system at $T=0$ upon an
increase in the number of electrons (decrease in the hole-doping level).
Until $\epsilon_m /2 > \epsilon_F$, only single electron quasiparticles
are present and the system behaves as an ordinary Fermi liquid. The
condition $\epsilon_m /2 = \epsilon_F$ determines the minimal
hole-doping level compatible with the state of a normal Fermi liquid.
Denote by $n_c$ the corresponding electron density $n$. Upon a further
decrease in the hole-doping level, all additional $n-n_c$ electrons
will pass into a Bose-Einstein (BE) condensate of pair quasiparticles (we
everywhere consider the case of small $n-n_c$ values, for which the
concentration of pairs is low and their interaction can be disregarded).
The system becomes a superconductor. The superconducting order parameter
represents the boson ground-state wave function $\psi\equiv \psi^{(x)}$
normalized by the condition $|\psi|^2=(n-n_c)/2$; wave function  transforms
in accordance with the $d_{x^2 - y^2}$ representation of the symmetry group of
\cuo\  plane.

It is important to note the following. In the system ground state (i.e.,
for complete filling of all fermion states with an energy smaller than
$\epsilon_F$), the uncertainty in the energy of a boson quasiparticle
with low excitation energy $\epsilon=k^2/2m$, arising due to its
collisions with single electron Landau quasiparticles, is proportional
to $\epsilon^2$. As in the conventional theory of Fermi liquid, this is
due, first, to the fact that the density of fermions
in the vicinity on the order of $\epsilon$ near $\epsilon_F$ with
which the given boson can collide due to energy conservation, is low.
Second, the statistical weight of the final states, to which fermionic
transitions are possible, is small. The probability of the boson decay
into two fermions per unit time is also small: as suggested in the
beginning of this Section, the boson must overcome a significant energy
barrier. Thus, the proposed picture of superconductivity in the vicinity
of maximal doping level remains valid even in the region of appreciable
densities of fermions, where the interaction between bosons and fermions
is significant. The critical electron density $n_c$ is determined from the
condition that the electron chemical potential is equal to half of the
minimal boson energy. In the general case, the latter is a functional of
the distribution function for single electron Landau quasiparticles.

In calculating the superconducting transition
temperature, the fermion distribution function may be considered as
corresponding to $T = 0$, since the temperature corrections (proportional
to $T^2$) to the thermodynamic functions of Fermi liquid are considerably
smaller than the corrections included below.

The density of uncondensed bosons at a finite temperature $T < T_c$ is
\begin{equation}
N'=\int\frac{2\pi k dk}{(2\pi\hbar)^2} \frac{1}{e^{\epsilon/T}-1}
=\frac{mT}{2\pi\hbar^2} \log \frac{T}{\tau}.
\end{equation}
The integral in Eq. (8) diverges at small $\epsilon$, so that it is cut
off at $\epsilon\sim\tau$, where $\tau$ is a small tunneling amplitude of
electrons in the direction perpendicular to the \cuo\ plane.

The excess number $n-n_c$ of
electrons in the system is equal to the doubled sum of $N'$ and number $N_0$
of bosons in the condensate. This leads to the dependence of the
superconducting transition temperature on the doping level for small
values of $n-n_c$
\begin{equation}
n-n_c = \frac{mT_c}{\pi\hbar^2} \log \frac{T_c}{\tau}
\end{equation}
and the number
of pairs in the condensate
\begin{equation}
N_0=\frac{n-n_c}{2}\left(1-\frac{T}{T_c} \frac{\log T/\tau}{\log T_c/\tau}\right),
\end{equation}
which determines the modulus of order parameter $|\psi|^2=N_0$ at finite
temperatures. The superconducting transition temperature defined by Eq.
(9) is quite high. To within the logarithmic term, this temperature is
on the order of one-electron tunneling amplitude $t$ at the
boundary of the applicability region (i.e., for $n-n_c \sim a^{-2}$). The
possibility that the superconducting transition temperature may have
such an order of magnitude was pointed out in the aforementioned paper
by Alexandrov and Kornilovitch \cite{lit6}.

The interaction of fermions with the BE condensate (effective
electron-electron interaction) that is described by the order parameter
$\psi$ creates an effective potential $\Delta_\k$ acting on fermions as
in conventional superconductors:
\begin{equation}
H_{int}=\sum\limits_\k (\Delta_\k c^+_{\k\uparrow} c^+_{-\k\downarrow} + \text{h.c.}).
\end{equation}
In view of the symmetry of $\psi$, we have
\begin{equation}
\Delta_\k=V(\hat{k}_x^2-\hat{k}_y^2)\psi,
\end{equation}
where $\hat{\k}=\k/|\k|$ and $V$ is invariant under the symmetry group.

Owing to this interaction, fermions in the superconducting state
considered acquire features typical of an ordinary superconductor with
the $d_{x^2 - y^2}$ symmetry.

\section{Normal state thermodynamics. Pseudogap}
The total number of pairs for $T < T_c$ is independent of temperature and equal
to $(n-n_c)/2$. The electron chemical potential for $T < T_c$ is also
temperature independent and equal to $\mu=\mu(n_c)=\epsilon_m(n_c)/2$
where $\epsilon_m=\epsilon_m(n)$ is the pair minimal energy which depends
on fermion density, as it is shown above.

For $T > T_c$, the fermion distribution function, as above, corresponds to
$T=0$, but with the temperature dependent chemical potential. The pair energy
spectrum is $E=\epsilon_m(\mu)+\epsilon$ where $\epsilon=k^2/2m$. The pair
density above $T_c$ is given by
\begin{equation}
N=
\int_0^{\infty}
  \frac{2\pi k dk}{(2\pi\hbar)^2}
  \frac{1}{e^{(\epsilon+\zeta)/T}-1} =
\frac{mT}{2\pi\hbar^2}\log\frac{1}{1-e^{-\zeta/T}}.
\end{equation}
The parameter $\zeta$ ($\zeta \gg \tau$) is defined by
\begin{equation}
\zeta=\frac{\partial \epsilon_m}{\partial \mu}\delta\mu -2\delta\mu
\end{equation}
where $\delta\mu = \mu-\mu(n_c)$.
With changing temperature, the total electron number conservation gives
\begin{equation}
n-n_c=2N+\frac{\partial n}{\partial \mu}\delta\mu.
\end{equation}
From the last equation, we find $\zeta=\zeta(T)$ and then all other quantities.

For $n > n_c$ and not too high temperature $T \ll T_c \log(T_c/\tau)$, the pair
density is determined by
\begin{equation}
\frac{N(T)-N(T_c)}{N(T_c)}=\frac{\partial n/\partial \mu}{2(2-\partial
\epsilon_m/\partial \mu)}Te^{-\Delta_p/T}
\end{equation}
where $N(T_c)=(n-n_c)/2$ and
\begin{equation}
\Delta_p = T_c \log \frac{T_c}{\tau} = \frac{\pi\hbar^2}{m}(n-n_c)
\end{equation}
is the pseudogap for $n > n_c$. As well as $T_c$, it is zero at the critical
value of the doping level $n=n_c$.
For higher doping level $n < n_c$, ($T_c =0$), we have
\begin{equation}
\label{NofT}
N(T)=\frac{mT}{2\pi\hbar^2}e^{-\Delta_p^\prime/T}
\end{equation}
where
\begin{equation}
\Delta_p^\prime =\left(2\frac{\partial \mu}{\partial n}-\frac{\partial
\epsilon_m}{\partial n}\right)(n_c -n)
\end{equation}
is the pseudogap for $n < n_c$. The equation \eqref{NofT} takes place in the low
temperature region $T \ll \Delta_p^\prime$. For $n < n_c$, the pseudogap
$\Delta_p^\prime$ is the gap in the energy spectrum of the pair quasiparticles.
For high temperatures $T \gg \Delta_p,\Delta_p^\prime$ (but $T \ll t$), the pair
density is a linear function  of temperature:
\begin{equation}
N(T) = \frac{z\partial n/\partial \mu}{2(2-\partial\epsilon_m /\partial \mu)}T
\end{equation}
where $z$ is the solution of the equation $\lambda z =e^{-z}$ with
\begin{equation}
\lambda =\frac{\pi\hbar^2}{m}\frac{\partial n/\partial
\mu}{2-{\partial\epsilon_m /\partial\mu}}.
\end{equation}

The entropy of pairs is determined by the equation
\begin{equation}
S(T)=\frac{m}{2\pi\hbar^2}
     \int_0^{\infty}d\epsilon\{(1+f)\log(1+f)-f\log f\}
\end{equation}
where $f=\left\{e^{(\epsilon+\zeta)/T}-1\right\}^{-1}$.
For $n>n_c$ in the low temperature region $T \ll \Delta_p$, we have
\begin{equation}
\frac{S(T)}{T}-\left(\frac{S}{T}\right)_{T=T_c} =
-\frac{m}{2\pi\hbar^2}\frac{\Delta_p}{T}e^{-\Delta_p/T}
\end{equation}
where
\begin{equation}
\left(\frac{S}{T}\right)_{T=T_c} = \frac{\pi m}{12\hbar^2}.
\end{equation}
$S(T)$ is almost linear in
$T$, with exponentially small deviations.
For $n<n_c$, the pair entropy is exponentially small at low temperatures
$T \ll \Delta_p^\prime$:
\begin{equation}
S(T)=\frac{mT}{2\pi\hbar^2}e^{-\Delta_p^\prime/T}.
\end{equation}
At high temperatures $T \gg \Delta_p,\Delta_p^\prime$, the entropy is
\begin{equation}
S(T)=\frac{m\sigma}{2\pi\hbar^2}T.
\end{equation}
The temperature independent factor $\sigma$ is determined by
\begin{equation}
\sigma ={\int_z^\infty\frac{xdx}{e^{x}-1}} - \lambda z^2.
\end{equation}
The entropy is again a linear function of temperature.

\section{Orbital paramagnetism of pairs}
In this Section, we show that the orbital motion of electrons inside the pairs
cause a peculiar paramagnetism. Let a pair is at rest as a whole. For singlet
pairs at $\k = 0$, the Hamiltonian (3),(4) can be written as a 2x2 matrix
\begin{equation}
H=4t
\begin{pmatrix}
0&1\\
1&0 
\end{pmatrix}
\equiv 4t\sigma_1,
\end{equation}
acting to the state vector $\psi$,
\begin{equation}
\psi = \psi^{(x)}
\begin{pmatrix}
1\\
0
\end{pmatrix} +
\psi^{(y)}
\begin{pmatrix}
0\\
1
\end{pmatrix},
\end{equation}
where $\psi^{(n)}$, $n=x,y$, are quantum amplitudes of two orientations of the
two-electron dumbbell, and $\sigma_1$ is a Pauli matrix.

In the $x$-state, coordinates of two electrons (with respect to the center of
gravity of the pair) are $x_1 =-a/2$, $y_1 = 0$ and $x_2 = a/2$, $y_2 = 0$,
respectively. In the $y$-state, we have $x_1=0$, $y_1=-a/2$ and $x_2=0$,
$y_2=a/2$. From this we find operators of coordinates for both electrons:
\begin{equation}
x_1 = -x_2 = -\frac{a}{2}
\begin{pmatrix}
1&0\\
0&0
\end{pmatrix},\quad
y_1 = -y_2 =-\frac{a}{2}
\begin{pmatrix}
0&0\\
0&1
\end{pmatrix}.
\end{equation}
The operators of velocities are determined by the commutators
\begin{equation}
\dot\r_{1,2} = \frac{i}{\hbar}\left[ H,\r_{1,2}\right].
\end{equation}
Simple calculation gives
\begin{equation}
\dot x_1=-\dot x_2=-\dot y_1=\dot
y_2=-\frac{2at}{\hbar}
\begin{pmatrix}
0&-i\\
i&0
\end{pmatrix}
\equiv -\frac{2at}{\hbar}\sigma_2.
\end{equation}
The operator of the pair magnetic moment which is directed along $z$-axis, is
\begin{equation}
\mu \equiv \mu_z =\frac{e}{2c}\sum_{1,2}(x\dot y -y\dot x) =
-\frac{eta^2}{\hbar c}\sigma_2
\end{equation}
where $e$ is the electron charge and $c$ is the velocity of light.

In the presence of an external magnetic field $B \equiv B_z$, the Hamiltonian of
the pair is
\begin{equation}
H=4t\sigma_1 - \mu B.
\end{equation}
The eigenvalues of energy are
\begin{equation}
\label{eigen}
E=\epsilon_0 \mp 4t\left[ 1+\left( \frac{ea^2}{4\hbar c}B \right)^2 \right]^{1/2}.
\end{equation}
In weak fields, the minimal energy is
\begin{equation}
E_{min} = \epsilon_0 -4t -t\frac{e^2 a^4}{8\hbar^2 c^2}B^2.
\end{equation}
The average magnetic moment of the pair is
\begin{equation}
\left< \mu \right> = -\frac{\partial E_{min}}{\partial B} = \alpha B,
\end{equation}
where
\begin{equation}
\alpha = \frac{e^2 a^4}{4\hbar^2 c^2}t = \frac{e^2 a^2}{4mc^2}
\end{equation}
is the pair paramagnetic polarizability.

We note that pairs with $\k = 0$ in the upper energy band (lower sign in
\eqref{eigen}) are diamagnetic.

The pair contribution to the paramagnetic susceptibility of 3D sample is
\begin{equation}
\chi = \frac{e^2 a^2}{4mc^2}N^{(3)}
\end{equation}
where $N^{(3)} = N(T)/L$ is 3D density of pairs and $N(T)$ is 2D density which
is determined by formulas (16), (18), and (20). Here $L$ is the distance between
neighboring \cuo\ planes.

Generally, we have three competing contributions to the magnetic
susceptibility: orbital paramagnetism of pairs considered above, Pauli spin
susceptibility of single electrons (pairs are singlet), and Landau diamagnetism
of single electrons and pairs. Spin susceptibility is isotropic. Orbital
paramagnetism and Landau diamagnetism are both strongly anisotropic (magnetic
moment is directed along $z$-axis independently of the direction of magnetic
field) because of 2D character of single electrons and pairs. However, Landau
diamagnetism, especially in 2D case, is very sensitive to inhomogeneities. For
example, it is easily suppressed by a localization of charge carriers. Orbital
paramagnetism is finite at zero velocity of a pair as a whole. So, it has to be
much more stable against inhomogeneities. We hope that orbital paramagnetism
can be experimentally separated from two other contributions to
the susceptibility.

\section*{Acknowledgments}
This study was supported by INTAS (grant no. 01686), CRDF (grant no.
RP1-2411-MO-02), Leverhulme Trust (grant no. S-00261-H), the Russian
Foundation for Basic Research (grant no. 03-02-16401), and the President
Program Supporting Leading Scientific Schools.

\end {document}